\documentclass[a4paper,notitlepage]{article}
\usepackage[latin1]{inputenc} 

\usepackage[T1]{fontenc} 

\usepackage{hyperref}
\usepackage{amsmath,epsfig,graphics,graphicx}

\title{Dynamic Load Balancing for PIC codes\\ using Eulerian/Lagrangian partitioning}

\author{
Marc Sauget\footnote{latu\ @\ unistra.fr, Universit\'{e} de Strasbourg, 7 rue Descartes, 67000 Strasbourg}%
  \and
  Guillaume Latu$\sp{\ast}$%
}

\date{December 2011}

\begin{document}


\newcommand{\myVoxel}{grid element}
\newcommand{\CmyVoxel}{Grid element}
\newcommand{\daywidth}{2.2 cm}

\maketitle \abstract{ This document presents an analysis of different
  load balance strategies for a Plasma physics code that models high
  energy particle beams with PIC method. A comparison of
  different load balancing algorithms is given: static or dynamic ones. 
   Lagrangian
   and Eulerian partitioning techniques have been investigated.  }

\section{Introduction}

Particle-In-Cell (PIC)  codes have  become an essential  tool for  the numerical
simulation of many physical phenomena involving charged particles, in particular
beam physics,  space and laboratory plasmas including  fusion plasmas. Genuinely
kinetic  phenomena can  be modelled  by  the Vlasov-Maxwell  equations which  are
discretized by  a PIC  method coupled  to a Maxwell  field solver.   Today's and
future  massively parallel supercomputers  allow to  envision the  simulation of
realistic problems involving complex geometries and multiple scales. However, in
order to achieve this efficiently new numerical methods need to be investigated.
This includes  the investigation of:  high order very accurate  Maxwell solvers,
the  use  of  hybrid grids  with  several  homogeneous  zones having  their  own
structured  or unstructured  mesh type  and size,  and a  fine analysis  of load
balancing issues. This paper improves a Finite Element Time Domain (FETD) solver
based  on high  order  Hcurl conforming  and  investigates the  coupling to  the
particles. We focus on the management of hybrid meshes that mixes structured and
unstructured  elements on  parallel  platform.  This  work  is the  result of  a
pluri-disciplinary researcher project HOUPIC~\cite{Jund07} bringing together the
IRMA's   team  and   the  LSIIT   team  of   university  of   Strasbourg,  INRIA
Sophia-Antipolis, the CEA, Paul  Scherrer Institut, IAG Stuttgart. 

\section{Context}

PIC method is widely used in the field of plasma simulations
(\cite{CarmonaC97},~\cite{Othmer02} and~\cite{walker90}). In such
software, a system dynamics of interacting charged particles is
influenced by the presence of external fields.  For the plasma we are
interested in, it is not computationally feasible to track the motion
of every physical particle. Thus, it is usual that simulations use
``super-particles'' that represents several physical
particles. Particles and fields that describe the physical problem are
represented by two computational objects, a set of ``super-particles''
and a mesh (structured or unstructured or hybrid 
that mixes structured and unstrucured) that discretizes the
spatial domain in order to stores the electric field $E$ and magnetic
field $B$. Each particle $P$, registers within an element of the
finite element mesh. Each particle knows the set of degrees of
freedom, the sources and the fields within the surrounding element,
denoted by ${FEM}_P$.  Classical PIC functionally decomposes into four
distinct tasks as depicted in figure~\ref{fig:PIC}
from~\cite{CarmonaC97}, namely:
\begin{figure}[h]
  \centering
  \includegraphics[width=8cm]{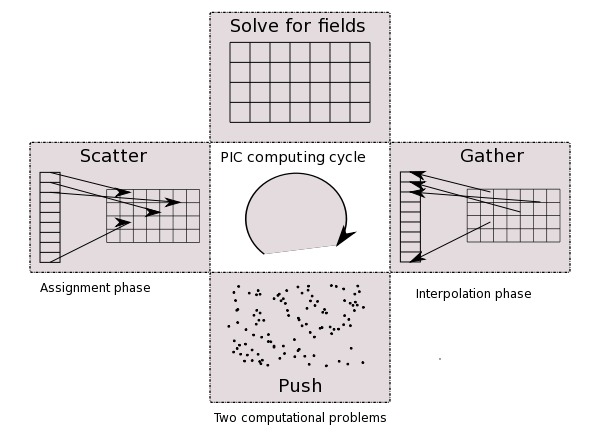}
  \caption{Phase of Particle-In-Cell algorithm}
  \label{fig:PIC}
\end{figure}

\begin{itemize}
\item[A.] {\textit{Assignment Phase:}} for each particle $P$, assign the charge and
  current density into ${FEM}_P$. This step aims at collecting the total charge and
  current density induced by the particle.
\item[B.] {\textit{Field  Solve Phase:}}  using the charge  and the  current density
  found in the previous phase  A, solve the Maxwell's equations (time integration) 
  through Finite Element Method to determine the electric and magnetic fields,
  $E$ and $B$. 
\item[C.] {\textit{Interpolation  Phase:}} for each  particle $P$, use the  value of
  $E$ and $B$ given by the Maxwell solver to interpolate the force value on $P$ at the
  particle's actual position.
  \item[D.]  {\textit{Push Phase:}}  update the  position of  each particle  under the
  influence of the force determined in previous phase C.
\end{itemize}

This four tasks are  performed at each time step of the  simulation run.  In the
set of 2D test  cases we are interested in, the \textit{Push}  step is often the more
costly part. In the sequel, we  will focus mainly on the parallelization and the
optimization of the last phase (denoted D). 

\subsection{Particles distributions methods}
\label{sec:PUSH}

Two methods are  generally used in Particles-in-Cell code  to distribute objects
over parallel machine.  These are  the Lagrangian and the Eulerian decomposition
methods. This  nomenclature is presented by Hoshino  et all~\cite{Hoshino88} and
presented with accuracy  by David W.  Walker in~\cite{walker90}.  The reason why
the two types of decompositions exist is  due to the fact that a PIC code should
consider  to  balance  the  load  of  the  mesh  firstly  or  of  the  particles
firstly. The  improvement of the  particle distribution over  processors reduces
the  cost of  the push  phase (D  step); whereas  a good  distribution  of finite
elements enhances the performance of the field solver (B step).
\\

The Eulerian method  manages particles motion in displacing  the particles among
processors  depending on  their  locations.  Each  processor  handles a  spatial
sub-domain.   Whenever particle changes  of spatial  sub-domain after  a motion,
this particle  is pushed away  from the processor  and is sent to  the processor
owning the new  sub-domain. A consequence of this method  is that each processor
manages all particles located inside a relatively small sub-domain.  This motion
of particles  between the  different processors constitutes  an overhead  in the
{\it Push} step.
\\

The Lagrangian method  allows each processor to track the  same set of particles
during all  the simulation on the  whole spatial domain.  Even  if the particles
are  initially in a  small area,  they can  travel far  away from  their first
location depending of fields they traverse.   In this case and after a while, it
is likely that one processor can manage particles dispatched almost everywhere
in the spatial  domain.  The main advantage of this method  over Eulerian one is
to  limit the  number of  communication  during the  \textit{Push} step  because
particles remain on the same  processor. Nevertheless, the execution time needed
by each processor to realize the \textit{Push} step are often larger than in the
Eulerian case because of a lack of cache phenomena. Hence, in the Eulerian case,
each  processor manages  particles  driven by  a  small set  of discrete  fields
included  in the  processor  sub-domain;  whereas in  the  Lagrangian case,  the
processor's set of particles can see all discrete fields of the spatial domain
during the push step. Thus, there are much more spatial locality in the usage of
field data  in the Eulerian  case than in  the Lagrangian one. The  cache memory
usage is better in Eulerian
decomposition as far as field data are concerned. \\

\subsection{Dynamic load balancing strategies} 

In  an Eulerian  mapping, there  are  two kinds  of method  concerning the  load
balancing strategy. The  first one, perhaps the most popular  one, uses a static
external       graph      partitioner       like      Metis       or      Scotch
(\cite{Karypis95metis,Pellegrini03nativemesh}).   This  external  tool performs
the partitioning of the mesh (and the  spatial domain) at the
beginning  of the  run, or  can also be  called off-line.   This processing can  have a
relative high cost compared to the  simulation runtime.  It is possible, but not
usual,  to  revise  the  partition  during  the  simulation  run, then using  the
partitioner multiple times per simulation. 
it may lead to a penalizing synchronization between processors.


We choose  to study an  alternate solution to  perform load balancing.  We would
like to  consider a dynamic  strategy to partition  the spatial domain  over the
processors that  is revised  during simulation.  The main idea  is to  adapt the
partition to a change of the macroscopic structure of particles.

In this case,  the partition tool is integrated directly  into the simulator. We
expect this  tool to be  cheap in  execution time and  to be used  frequently to
adapt the  partition to the dynamics  of particles.  We will  focus hereafter on
two methods implementing that kind of load balance strategy.  The first one is a
global  partitioning  method that  considers  the  whole  space domain  and  the
computation cost (via  a cost function) on each spatial unit.  It is denoted URB
(Unbalanced Recursive  Bisection) in the  literature. The second method  we will
look at,  is known  as \mbox{ORB-H}.  This method prevents  the synchronization  of all
processors during  a dynamic  mapping phase.  This  feature could be  great when
working on  large scale platform, compared  to the previous URB  algorithm as we
shall see afterward.  In the two cases,  URB and ORB-H, we choose to give to the
partitioner  an input  rectangular shaped  \textit{\myVoxel{}s} (for  a domain  in two
dimensions). That means that  each uncuttable rectangular \textit{\myVoxel{}} encompasses
a set of finite elements and particles.  We determine the responsible \myVoxel{}
for  each element  depending  on the  barycenter  location of  the element.  The
partitioner does  not see finite elements  nor particles but only  the number of
particles and finite  elements encapsulated within a \myVoxel{}.  By the way, we
avoid  the  difficulty of  managing  complex  data-structure  needed for  finite
elements  and  we avoid  the  computation  of a  partition  of  a general  mesh.
Furthermore,  this approach  facilitates  the  tuning of  a  partition based  on
rectangular  sub-domains (see \cite{Berger87}).  We will  evaluate this  kind of
partitioning method.

As  we  already  said,  this  study  focuses  on  the  parallel  management  and
distribution  of  particles  in a  PIC  code  in  order  to mainly  enhance  the
performance  of the \textit{Push},  \textit{Gather}, \textit{Scatter}  steps and 
not on the \textit{Field solver}. As a perspective, we
wish to extend  our results to the load balance of  all steps of the code
including the  field solver. In order  to do that,  we will have to  improve the
cost  function given  to the  partitioning algorithm  to take  into  account the
number of finite elements per \myVoxel{}.  Suppose we want to consider test cases
that includes  a costly Maxwell  solver part; a  small modification of  the cost
function will allow us to tackle this new configuration.

\subsection{\CmyVoxel{}s}

\begin{figure}[h]
  \centering
  \includegraphics[width=4cm]{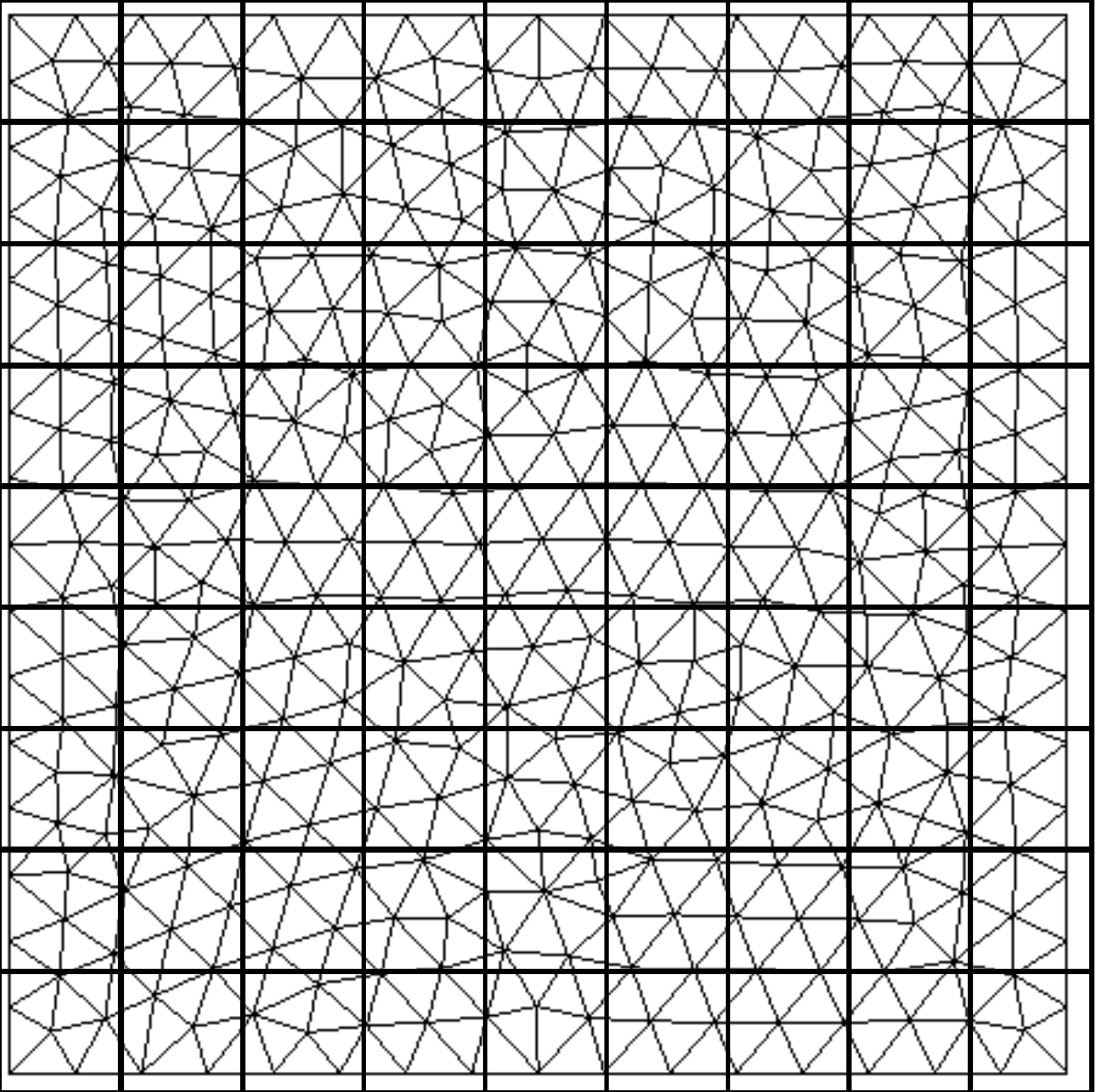}
  \caption{The initial \myVoxel{} cutting}
  \label{fig:grid}
\end{figure}

The figure~\ref{fig:grid} illustrates a sketched  view of our spatial domain.  A
regular grid divides the domain  in \myVoxel{}.  Each \myVoxel~has a rectangular
shape and a  fixed size defined as a parameter of  the simulation.  A \myVoxel{}
stores a set of finite  elements (mesh information and field solver information)
and particles.  The first step of  our simulation tool assigns each particle and
finite  element information  to one  \myVoxel{}.   We store  also the  belonging
relation  existing between particles  and \myVoxel{}.  This relation  is updated
after each particle motion.
The  \myVoxel{} sub-grid  offers a  way to  simplify several  mechanisms. First,
the \myVoxel{}s  allow  the parallel  simulation  to  migrate  finite elements  and
particles  easily from  one processor  to the  other (from  the  programming and
software engineering  point of  view).  The penalty  in term of  performance and
partitioning quality is  low if the number of  \myVoxel{}s per processor remains
large.   Second, the  localization  of particles  is  trivial for  this kind  of
Cartesian sub-griding, which is a big advantage for a PIC code. Furthermore, one
could derive a  simple condition to detect when a  particle leaves the processor
sub-domain.

\subsection{Initial partitioning}

If the global space  domain is a rectangle, one of the  simplest mapping one can
think  of,  is  a   homogeneous  decomposition  of  rectangular  sub-domains  on
processors.  For the time being, we  do not consider the number of particles per
\myVoxel{} as a constraint  but  simply   distribute   the  same   number   of  \myVoxel{}s   per
processor. Here, we do not care  about the cost function but propose a reference
mapping for comparison purpose.


\begin{figure}[htpb]
  \centering
{\footnotesize
\begin{verbatim}
TYPE SUBDOMAIN
  INTEGER   pid                  ! processor id
  REAL*8    xmin,xmax,ymin,ymax  ! rectangle definition
  INTEGER   height               ! height of the node in the tree
END TYPE SUBDOMAIN
\end{verbatim}
}
  \caption{Data type used in the parallel domain decomposition}
  \label{verb:URBCell}
\end{figure}

In several algorithm, the parallel domain decomposition is performed in building
a tree.   The data structure  that constitutes  a node of  the tree is  shown in
figure~\ref{verb:URBCell}.   Each   sub-domain  is  associated   with  a  single
processor.  The field  $pid$  contains the  processor  identity.  The  following
fields define  a rectangular area which  corresponds to the  sub-domain given to
the processor. The integer $height$ stores the height of the cell in the tree of
the hierarchical domain decomposition.\newpage

\section{Different strategies for load balancing}

Parallelisation of a PIC code  relies to a large extent on the load balancing 
chosen
and  the partitioning  scheme  used.  The partitioning  scheme  can be  either
hierarchical or non-hierarchical.

In  a   hierarchical  partitioning,   the  mapping  is   derived  from   a  tree
representation  of sub-domains. One  can directly  use the  tree to  adapt the
mapping to minor changes of the load.

Hereafter, we will look at a few Recursive Coordinate Bisection (RCB) algorithms
and  an  incremental  partitioning  solution  known as  the  dimension  exchange
algorithm.

\subsection{Simple geometric partitioning algorithm}

The first algorithm we look at to perform the load balancing uses an improved
version of  the  algorithm presented  by  Berger  and
Bokhari~\cite{Berger87}, known  as Recursive  Coordinate  Bisection.   
The RCB algorithm  divides  each  dimension
alternatively  in  half   part  to  obtain  two  sub-domains having
approximately the  same cost (using the user-defined cost
function). The two halves are then  further divided by applying the
same process 
recursively.  Finally,  this  algorithm  provides  a  simple  geometric
decomposition for a number of processors equal to a power of two.

\begin{figure}[h]
  \centering
  \includegraphics[width=8cm]{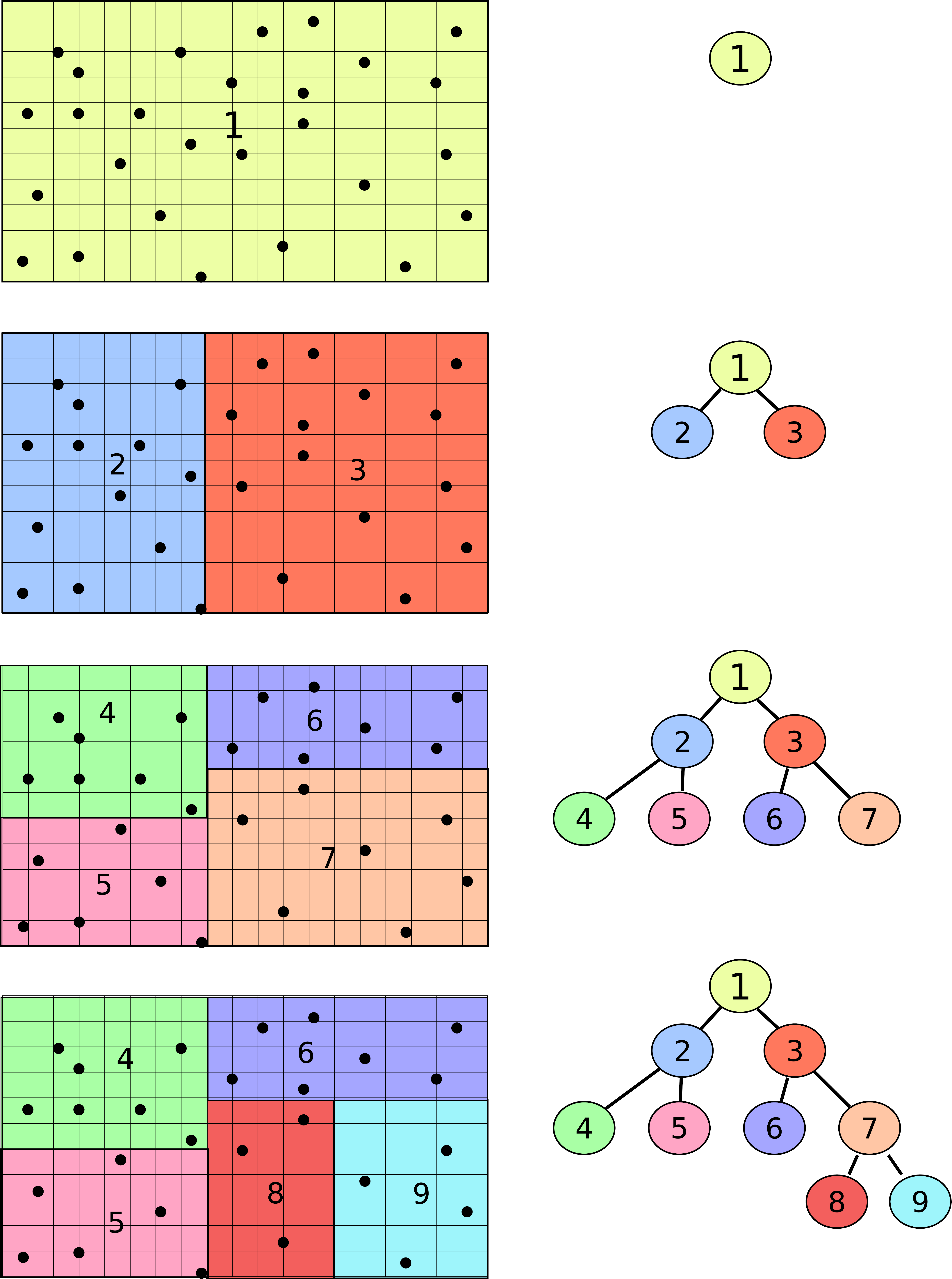}
  \caption{The URB partitioning strategy}
  \label{fig:URB}
\end{figure}

This partitioning  algorithm does not pay  attention to the aspect  ratio of the
resulting sub-domains. If  there is a large variation  between dimensions of the
domain, one of  the two sub-domain boundaries can be very  large compared to the
other.  Since the communication volume is  related to the size of the sub-domain
boundary (A and C phases),  large aspect ratios are undesirable.  The Unbalanced
Recursive Bisection (URB) is an  improvement of the previous partitioning method
proposed by Jones  and Plassmann~\cite{Jones}.  They propose to  try to minimize
the aspect ratio and suppress the  constraint of using power of two sub-domains.
The main  idea is to  adjust the  sub-domains size in  respect to the estimated cost.


In  our PIC  code,  we do not  manage  directly the  finite  elements but  simply
rectangular \myVoxel{}s as  we described  previously. We  used the  load per  cell to
dispatch them on the different nodes.

The figure~\ref{fig:URB} illustrates an  URB decomposition process.  Firstly, we
begin  with a  big cell  containing  all the  processors and  the whole  spatial
domain. At the start of this algorithm, the first cell contains all \myVoxel{}s.
Recursively, we  divides each remaining  cell of size ($n$,$m$)  \myVoxel{} into
two  parts in  assigning  to each  new cell  a  distinct set  of processors  and
distinct sub-domains. To divide the set of \myVoxel{}s, we use the cost function
to find the right cut.

We  obtain  two cells  that  have  respectively the  size  of  ($n-q$, $m$)  and
($q$,$m$) (if the cut  is in the first dimension).  The choice  of the q-size is
determined using  a simple method like  the dichotomy to  share the load with a fair and
equitable strategy between the two cells. In  a second step, the created cells are divided
again  in the  other dimension.   We  choose the  algorithm to  perform the  cut
alternatively in  each dimension  (\mbox{x or y}).   This last property  of alternation
allows  one to  obtain  sub-domains  with quite small  perimeters.   This choice  is
worthwhile   because    the   volume   of    communication   between   different
processors~\cite{Berger87} is  proportional to the  sub-domain perimeters during
scatter and gather phases.  At the end of the recursion, the whole domain is split
in balanced load parts. A tree  structure is generated that stores each cut and
finally the data structure of figure \ref{verb:URBCell} in the leaves.

In our  code, particles move  at each iteration  of the run  between processors.
Thereof  it  is necessary  to  regularly  apply  the partitioning  algorithm  to
maintain  a  good  load balance.   From  one  iteration  to the  next,  the
partitioning algorithm we just described, can lead to very different mappings.
Indeed, the algorithm does not try to keep or adjust the previous partition, but
compute from scratch another partition.  However, a big change of the partition
can provoke  a massive migration  of objects between processors  and can
induce ta degradation of application performance. We will  address the problem  in the next
section.

\subsubsection{URB limited migration}

\begin{figure}[h]
  \centering
  \includegraphics[width=8cm]{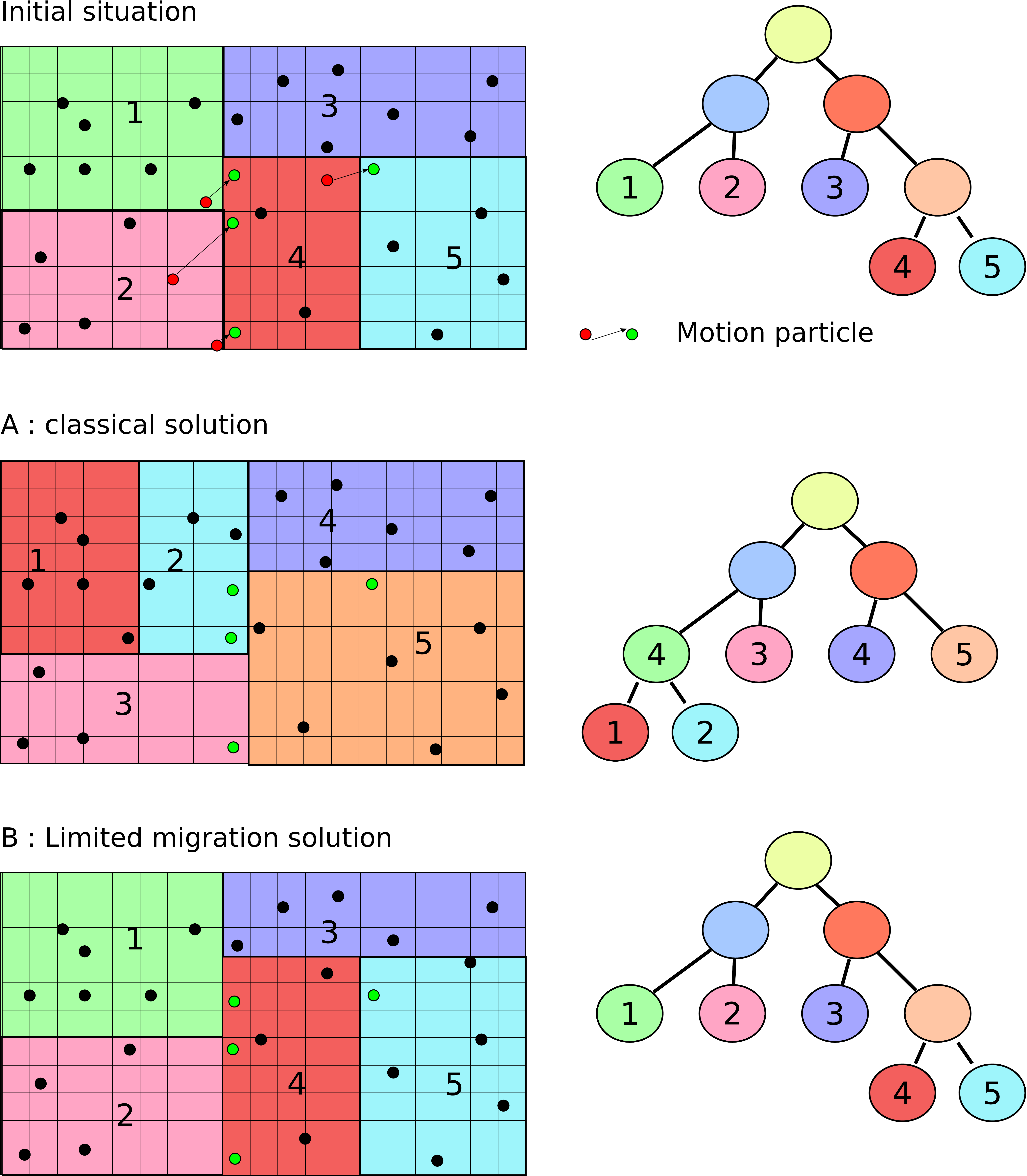}
  \caption{Illustration of the limited migration URB}
  \label{fig:URB3}
\end{figure}

We have designed  a simple solution to constraint the  load balancing process to
use quite  similar decomposition  from one iteration  to the other.   Remind you
that the tree structure is built  during the partitioning process. We propose to
modify a little  bit the existing tree instead of building  a new tree.  Indeed,
we adapt the  algorithm to only authorize load  exchange between cells belonging
to the same  branch of the tree, or,  up to a maximum level  of the partitioning
tree.  We  set that only one frontier  in each sub-domain is authorized to move during the
URB algorithm.  Consequently, we stop the  risk of a global  modification of the
partition,  and  then  we  limit  the  migration  of  data  between  processors.
Nevertheless,  if particle  motions are  large enough,  and if  the  global load
balance  becomes not  so  good, this  version  of load  balancing that  preserve
locality could be inadequate.  If ever this unusual configuration happens, it
would  be  useful  to call  the  standard  URB  load  balancing to  correct  the
imbalance.

The difference between  the classical URB and the modified  version is shown in
the figure~\ref{fig:URB3}. The $A$  case corresponds the classical URB algorithm
applied after a migration  of the particles.  One can see a  large change of the
domain decomposition.  Many \myVoxel{}s  and associated particles have to migrate
on their new  processors.  In opposite, the limited  migration URB restrains the
modifications of the  tree to happen in nodes higher than  the third level.  The
result is that the new tree is close to the old one. Only the boundaries of three
sub-domains are moving.\\

Finally,  this new  version  of the  algorithm  fills our  first objectives.   A
remaining  problem  is about  the  synchronization  still  required between  all
processors  during the  partitioning algorithm.   It is  a bad  point  for large
parallel platforms.   The tree is unique  for all processors and  obtained with the knowing
of  the loads computed  in each \myVoxel{}.   This adds a  communication step
that can penalize  the global performance of the  application.  In the sequel,
we propose  a \textit{local} algorithm  that computes dynamically  a partitioning
avoiding global synchronization.
\subsection{A diffusion algorithm for the load balancing}
\label{sec:DIM}

A communication  bottleneck is a major  drawback for a parallel algorithm.  The
previous  algorithms need  a  gather step  to  share information  and require  a
synchronization. All processors are waiting  to each other, and in worse cases,
the time loss could be significant.  A simple way to overcome this bottleneck is to
avoid  the global  communication and  replace  it with  communications only  towards
neighboring  processors.  For  load balancing  purpose, Cybenko~\cite{Cybenko89}
described an incremental algorithm that uses the first order diffusion equation.
At  each  simulation  iteration,  this  diffusion algorithm  proceeds  to  local
exchanges of load between closer processors. It converges asymptotically (for an
infinite  number   of  iterations)  towards   a  uniform  load   balance.  Many
improvements  of  this  algorithm  have  been published  and  we  highlight  the
following papers\,\cite{JeannotV06,Watts98} for the practical description given.

\begin{eqnarray}
  w^{t+1}_i = w^{t}_i +  \sum_j{\alpha_{i,j} (w^{t}_j - w^{t}_i)} + \eta^{t+1}_i
  -c_i
  \label{equ1}
\end{eqnarray}
The algorithm  is based on the  diffusion equation (\ref{equ1}) that  is able to
estimate  the diffusion  of  the load  between  processors in  the framework  of
dynamic load balancing.   In the previous formula, $w^{t}_i$  corresponds to the
load  of processor  $i$ at  time $t$.  $\eta^{t+1}_i$ is  the load  who  will be
created  on the processor  $i$ at  time $t$.  The consumed  load by  a processor
during  an iteration is  $c_i$. In  our application,  the $c_\star$  constant is
always zero  because the work to  perform is the  same at each time  step.  Work
does not diminish along time dimension.

In this  algorithm, all processors can  exchange load at  each iteration. But,
sometimes,  it is  not possible  to exchange  load easily  without  paying large
communication costs.  At the utmost, we  can even choose to impose that a node
$i$ can only exchange  load with one node $j$ at an  iteration $t$ in order to
reduce   amount    of   communication.     This   solution   is    proposed   by
Cybenko~\cite{Cybenko89} and  this algorithm  is known as  the \textit{diffusion
  exchange} algorithm.  This algorithm use processors  two by two  to obtain the
asymptotic stability state\,\cite{JeannotV06}. \\


For our simulation, we fix the parameter $\alpha_{*,*}$ to $1/2$ uniformly and 
$\eta^{*}_*$ to $0$. The formula~\ref{equ1} becomes :

\begin{eqnarray}
  w^{t+1}_i =  w^{t}_i  + \frac{1}{2} (w^{t}_j-w^{t}_i)
  \label{equ2}
\end{eqnarray}

\begin{figure}[h]
  \centering
  \includegraphics[width=5cm]{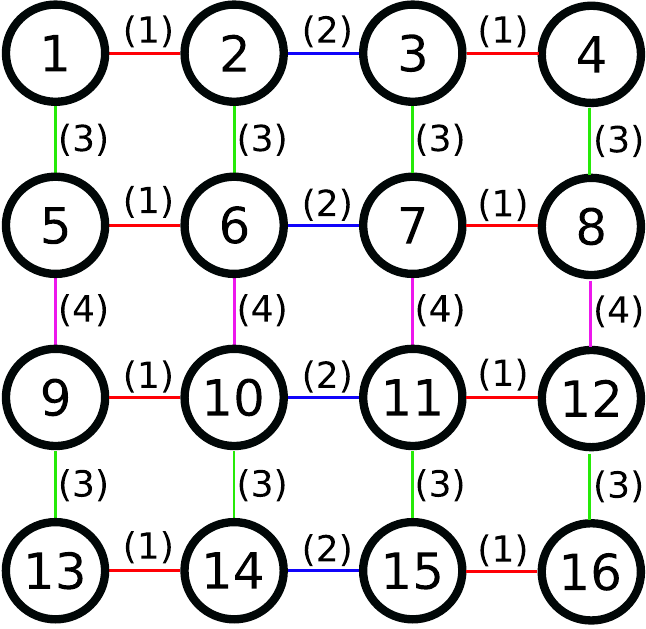}
  \caption{Coloured grid 4x4}
  \label{fig:dimExC}
\end{figure}

The scheme of figure~\ref{fig:dimExC} shows a 4x4  grid of processors.  Capability of load
exchange is sketched by  a link between two nodes. Using a  colored graph, it is
easy to deduce  at the iteration $t$ which processors  exchange load. The figure
\ref{fig:dimExC} describes a domain in two dimensions.


In this figure, we can classify the processors into two different sets.  The first
set, composed by the processor place on  the middle of the domain, have all of
its communication  channels with  the other processors.   In the context  of the
figure~\ref{fig:dimExC}, this  class contains the processors 6,7,10  and 11.  For
example,  the  processor  6 can  exchange  data  with  the processor  5,  when
$mod(it_{number},4) = 1$,  with the processor 7, when  $mod(it_{number},4) = 2$,
and with  the processor 2 and  10 in the  corresponding case. The other  case of
processor (the  number 1,2,3,4,5,8,9,12,13,14,15 and 16) has  the same behavior
as the  first class usually.   But, in some cases,  as this
processor  have  not  all of  the  channels  of  communication, the  results  of
$mod(it_{number},4)$ give  a choice of channels not available on  the processor
(like the channel~1 for the processor  1). In this case, this processor does not
participate at the load balancing step for this iteration.

\begin{figure}[h]
  \centering
  \includegraphics[width=8cm]{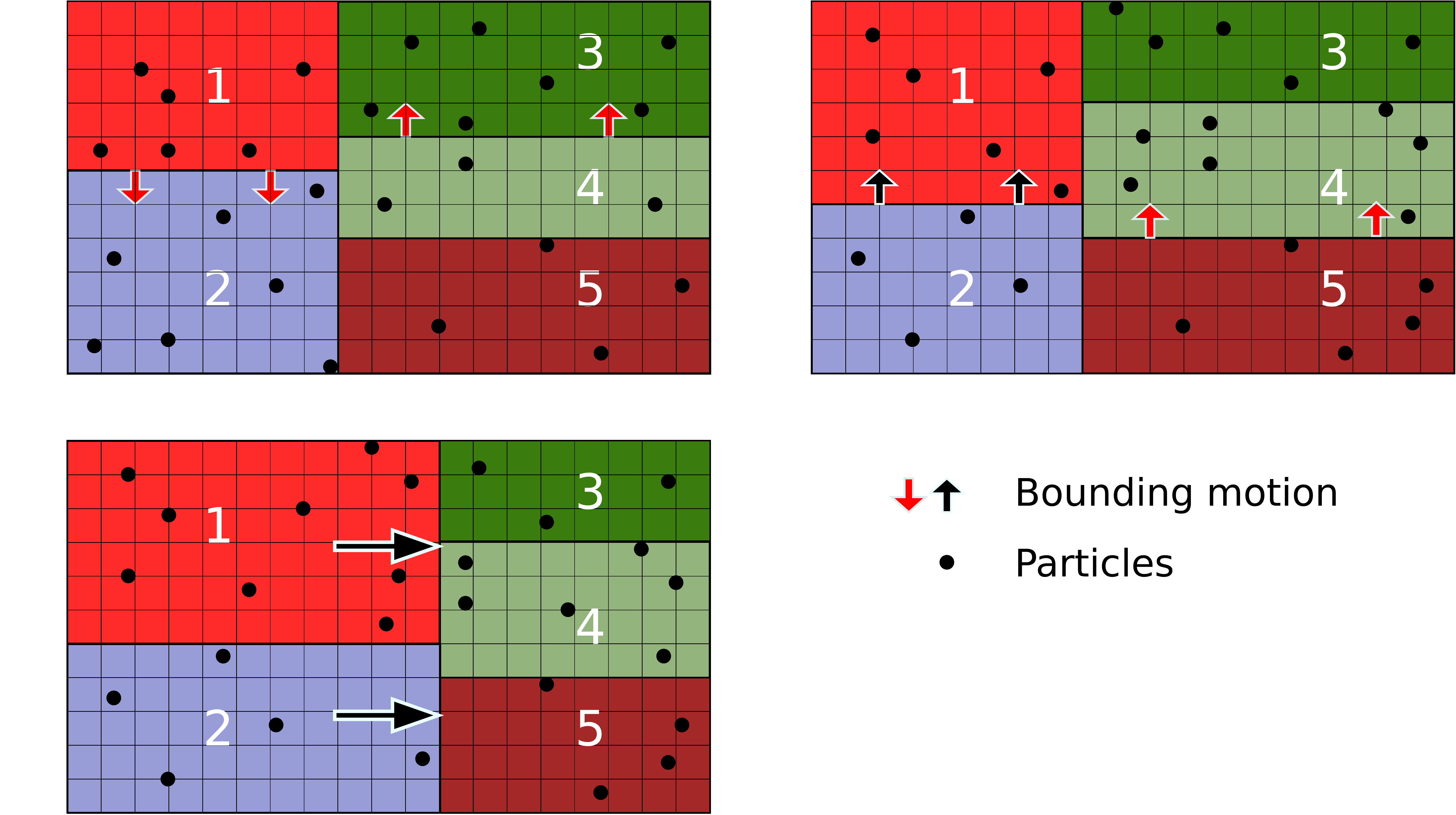}
  \caption{The 
  -H partitioning strategy}
  \label{fig:ORBH}
\end{figure}

A specific design is needed to take into account the distributed feature of this
algorithm. As we  already have described in the previous  section, the first partitioning
strategy  choice  is  not  very   useful  to  exchange  load  between  different
neighbors. The only allowed choice is the cell in the same level or in the same
branch on  the partitioning tree.  As  we want to have more  possibilities to exchange
load, we  have preferred to  use an another  strategy of partitioning:  the \mbox{ORB-H}
strategy   proposed  by   M.   Campbell,   A.    Carmona  and   W.   Walker   in
\cite{Carmona90,CarmonaC97}.   This strategy  divides  each different  dimension
separately and  does not need the  use of another data  structure to store  the global
scheme  of  the  partitioning.   The  figure \ref{fig:ORBH}  presents  the  ORB-H
decomposition for the example already presented in \ref{fig:URB}.

The scheme \ref{fig:ORBH}  shows that all the borders are  not independent as for
the  URB partitioning.   The exchange  of  load cannot  be realized  in all  the
dimensions at a time and only  the last partitioned dimension  can be easily used
for the load balancing process (in 1D).  In practice, for our
application, each node has the choice between at most  two nodes at each iteration to proceed to
a load exchange.

The work domain is the same as in  the previous case. The domain is 2D and use a
first regular  decomposition of \myVoxel{}s.  The initializing  process is realized
in two steps.  First, the  algorithm divides the domain in equally sized parts
using  only one  dimension load  balancing. The  choice of  the  first dimension
dividing the global  domain is important and must be  chosen adequately with the
particle  motions.  At  the next  iteration, the  process divides  the second
dimension in equivalent load parts.

During  the  global process,  particles  moves in space and  a
situation  of load  imbalance  could  appear. In  the  previous algorithm,  this
detection must  be solved by  a global process  of load balancing  involving all
nodes. In  the case of a distributed  algorithm, only the node  affected by this
situation try to solve it with its closest neighbors. \\

The  diffusion exchange  algorithm uses  the following  principles :
\begin{itemize}
\item  one node  can  exchange its  load with  only  one neighbor  in the  same
  sub iteration of load balancing,
\item   the   volume  of   the   load  exchange   is   given   by  the   previous
  formula~\ref{equ1}.   In our  case, the  volume  correspond to  the number  of
  \myVoxel{}s that can migrate between two nodes.
\end{itemize}

With  the  simple scheme  (figure~\ref{fig:ORBH}),  the  exchange of  load  is
possible between the  node 1 and 2, 3  and~4.  The node 5 does  not perform load
balancing  step  at  this  first  iteration.   At the  next  iteration  of  load
balancing, the exchange of data uses the same configuration for the first column,
but for  the second, only  the 4  and 5 exchange  load. And less  frequently the
column of the cells 1 and 2 could exchange load with column of the
cells 3,4 and 5.  We have implemented  a predictive method to do the choice of a
couple of nodes exchanging their load using the parity of the node.  The implementation
of this  algorithm uses the same  structure than the URB  algorithm.

\section{Results}

\subsection{Numerical experiments}

We  have tested our  algorithms  in many  situations  considering few  parameters
variation. To limit the  number of cases to run, we have  fixed a limited number
of  parameters as  the  number of  particles  (eighty millions),  the number  of
iteration to  be computed  (one thousand) and  the type  of test case.   We have
studied the case of the  "Two stream instability"~\cite{TSI01} in two dimensions.
This case is a periodic case, so the number of particles, equivalent to the cost
function load for us in this study, is the same along the evaluation, there
is no loss of particles.

\begin{figure}[h]
  \centering
  \includegraphics[width=3cm]{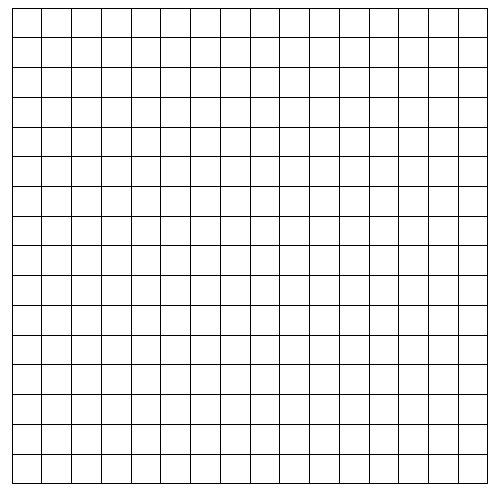}
  \includegraphics[width=3cm]{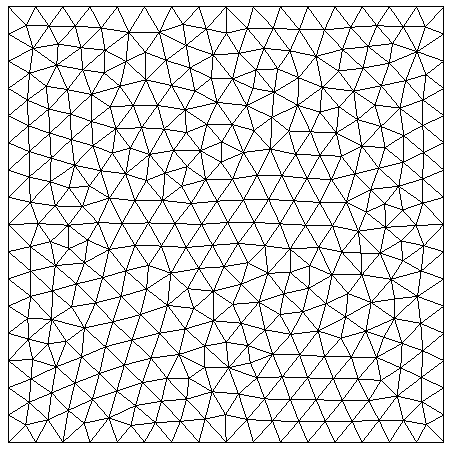}
  \includegraphics[width=3cm]{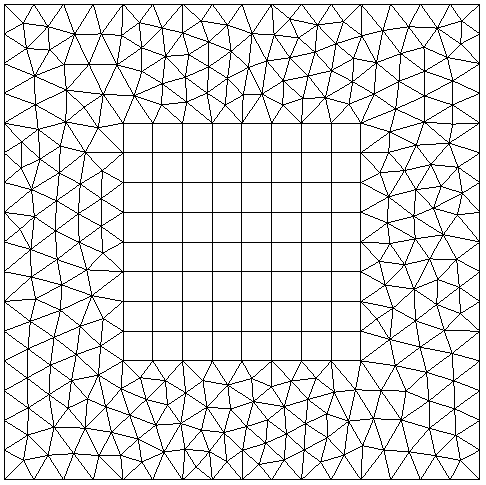}
  \caption{Regular mesh (left), unstructured mesh (middle) and hybrid mesh (right) }
  \label{fig:MAIL}
\end{figure}

We have  tested our simulation  using two distinct types of mesh,  quadrangles and
unstructured  mesh.  A  view  of this  two types  of  mesh are  shown in  the
figure~\ref{fig:MAIL}.

Parameters  used to study  the efficient  of different  types of  load balancing
algorithms  are the  following. We  have tested  the variation  of the  number of
vertices and  edges of the used  mesh,  and the  type of
mesh (triangles or quads). We have used  the dedicated high-performance facilities of the
University of Strasbourg.  This cluster  has 512 processors dispatched within in
64 bi-processor nodes (Intel Opteron 2384).

\subsection{Dimensionnal study for the two kinds of mesh}

A first comparison  has the objective to show the  behavior of several algorithms
depending on the mesh size.  This  test uses  the common
setting with a grid of  $4 \times 8$ processors. The two evolving
parameters are the type of mesh, and its size ($16 \times
16$, up to  $128 \times  128$).

\begin{table}[htpb]
  \footnotesize
  \centering
  \begin{tabular}[h]{| r|r | r |c | r|r |c|}
    \hline
    & \multicolumn{2} {|c|} {Regular mesh}  & \multicolumn{2} {|c|} {Unstructured  mesh}  \\
    \hline
    & \multicolumn{1}{ |c|} {Nb of quads} & time     &  \multicolumn{1}{|c|}{ Nb of triangles } & time  \\
    \hline
    16x16   &   256 & 7847 s  &   696 & 7665 s\\
    32x32   &  1024 & 8076 s  &  2730 & 7575 s\\
    64x64   &  4096 & 8417 s  & 10574 & 7830 s\\
    128x128 & 16384 & 8041 s  & 35930 & 8568 s\\
    \hline
  \end{tabular}
  \caption{Dimensional study for the two kinds of mesh} 
  \label{tab:DSM}
\end{table}

The results and the sizes of the meshes
are presented in the table~\ref{tab:DSM}. It shows the very good behaviour of our
implementation depending on the evolution of  the size (i.e. increasing the mesh size does not induce
 an increase of the execution time which is mainly prescribed by the number of particles).   Indeed, the
global time  of a  simulation is relatively  independent of the  mesh size
because the time used in the {\it{field solve phase}} (see figure~\ref{fig:PIC})
is small, around five to ten percent of the global time of an iteration.

\subsection{Comparison between different methods for particles  management}

\begin{table}[htpb]
  \footnotesize
  \centering
  \begin{tabular}[h]{| c |r | r|}
    \hline
    &  \multicolumn{1} {|c|}{ Lagrangian }&  \multicolumn{1} {|c|}{  Eulerian }\\
    &  \multicolumn{1} {|c|}{ (P1) }    &  \multicolumn{1} {|c|}{  (P3)} \\
    \hline
    static  (1)      & 21175s  & 8325 s   \\
    static* (2)      & 22519s  & 8948 s  \\
    URB                 &  9212s  & 8684 s \\
    URB-limited      &  9682s  & 8376 s \\
    ORB-H             &  9174s  & 8568 s \\
    \hline
  \end{tabular}
  \caption{Global evaluation for particles push methods} 
  \label{tab:PUSH}
\end{table}

\begin{figure}[h]
  \centering
  \includegraphics[width=3.9cm]{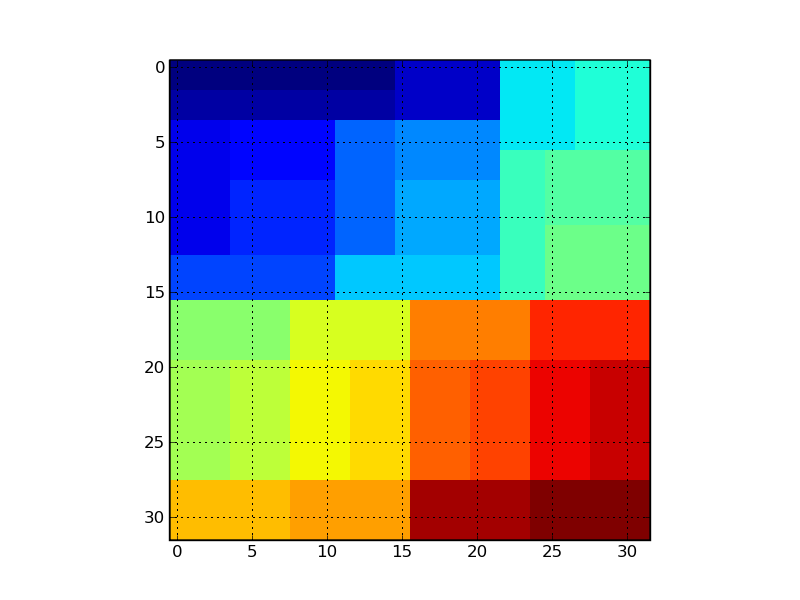}
  \includegraphics[width=3.9cm]{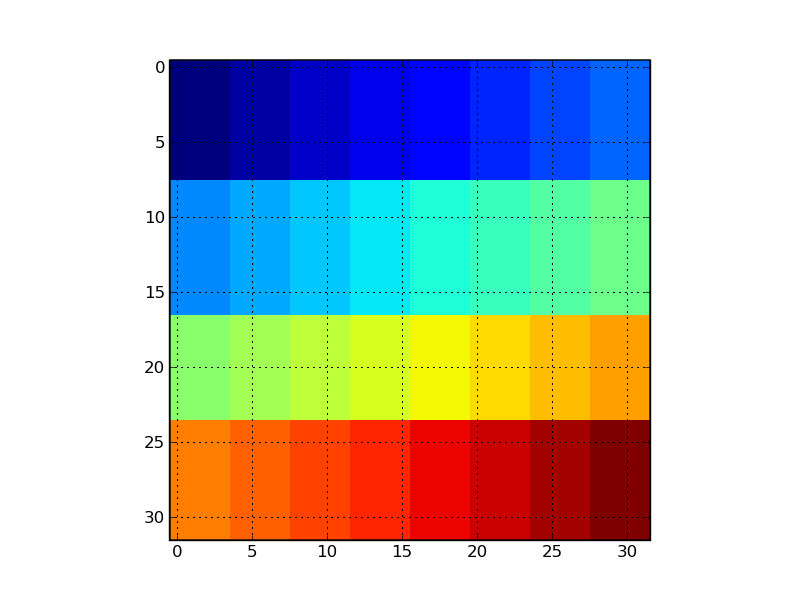}
  \caption{Comparison between URB (on the left) and ORB-H (on the right)  dynamic load balancing process. Each color represents  a processor's domain.}
  \label{fig:RES_ORBH}
\end{figure}

This comparison aims  at showing the difference between the  two methods (Lagrangian and Eulerian) depending on the
partitioning algorithm.  As it is shown in the section~\ref{sec:PUSH}, the
main difference  between these two methods is  that the spatial locality effect in the cache
can be favored. The particles that are close to each other use the same data of the elements which can favor cache effects. There is
also a benefit brought by  avoiding particle migrations.   This test uses the two
partitioning methods described previously and illustrated in the
figure~\ref{fig:RES_ORBH}.

To  show  main  differences between  these  two  methods,  we have  compared  the
difference on  the global time  for one simulation  and the time needed  for one
iteration during the simulation for  each load balancing algorithm for static and
dynamic  versions. The  table~\ref{tab:PUSH} presents  the results  of  the global
computation time for each case of  the study. The main finding
is  about  the differences between  the  static and  the dynamic  load
balancing. In  the Eulerian  case, the difference  is not existent  because this
method  of particles motion  allow the  localized push  of particles  within each
processor. For the Lagrangian method,  the static method are
very penalizing for the particles motion cost. In this case, each processor keep the same 
set of particles during
the whole simulation  run. After  many iteration of  calculus, each  processor manages
particles  dispersed over the whole  simulation  domain. This  context of  calculus
does not facilitate the optimization of the execution and induces an increase by a factor
two at least on the global time of the simulation.

\begin{figure}[h]
  \centering
  \includegraphics[width=8cm]{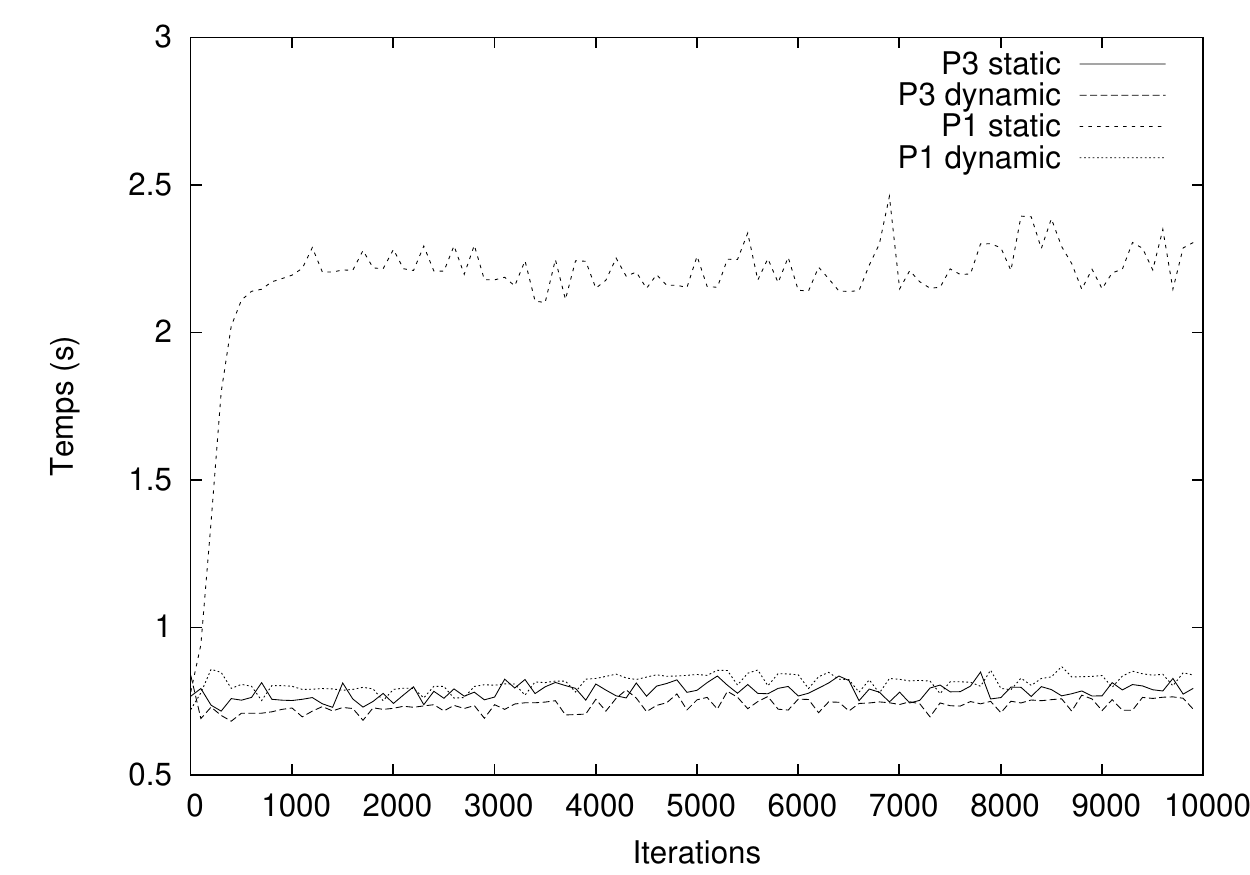}
  \caption{Importance of the method for particle motions}
  \label{fig:PUSH}
\end{figure}

The figure~\ref{fig:PUSH}  presents the computation  time for each  iteration in
the previous context. In the two cases, static and dynamic, the Eulerian method (P3) is
more efficient compared to the Lagrangian  method (P1). This optimization is clearly visible for
the static case.  Indeed, in the dynamic  case, particles are re-localized and re-distributed at each load
balancing step.

The improvement brought by the Eulerian dynamic method is
around 15\%  on the execution time  compared to Eulerian static.

\subsection{Comparison  between static and  dynamic load  balancing for  the two types of mesh }

\begin{table}[htpb]
  \footnotesize
  \centering
  \begin{tabular}[h]{ |l|l|l|l|l|}
    \hline
                 &static  &  URB         & ORB-H  \\
                 &  URB   &``limited'' &   \\
    \hline
    Regular      & 7921 s & 8250 s       & 8041 s \\
    \hline
    Unstructured & 8325 s & 8376 s       & 8568 s \\
    \hline
    Hybrid       & 7651 s & 7907 s       &  8328 s \\
    \hline
  \end{tabular}
  \caption{Global evaluation for the two families of mesh (using a $128\times128$ mesh)} 
  \label{tab:GETF}
 
\end{table}
This test  shows the difference  of performance of  our simulator using  the two
types of mesh.  This benchmark uses  the common configuration like all the previous
benchmarks.   We  have  choose  to  do  this  bench   with  a mesh size  of
$128\times128$  and we  have using  the Eulerian  distribution method  to manage
particles.  The   regular  mesh  is   compound  of  16641  vertices   and  16384
quads. Respectively,  the unstructured mesh is compound  of 18222 vertices
and 35930 triangles.  The results  are detailed in the table~\ref{tab:GETF}.  In
conclusion to  this, we can say  that the difference of  performance is
existent but  not very large.  Indeed, the  maximum difference  between two
execution times is  about ten percent.  This small difference of  performance between
these two types  of mesh exists because  most of the time  consumed
within one iteration is consumed by the push particles function (neighborhood management and communications represents less than 10 percents).



\section*{Conclusion}


We  have  studied  in this  work  several  algorithms  of  load balancing  for a PIC code using  
static or dynamic partitioning.  We  have shown that the dynamic management is a key point
 to optimize the overall performance  of the application. But,
we  have not shown  a  large  difference between  the  load balancing  algorithms under investigation.
Indeed, the measured times remain close to each other for the considered algorithms.   The localization  of the
exchange with the limited version of the URB partitioning or using the ORB-H have
not led to the  improvement we hoped. Two facts can explain  this.  The first is
about  our  developments of parallel algorithms. There  are  some points  of synchronization 
we should remove that
are killing performance. The second point
is that the use cases which are  maybe not well designed to evaluate
the different load balancing algorithms (not enough particles moving around).  
The perspectives are, first,  to
find a  another use case to  evaluated with accuracy the  load balancing algorithms
and second, to study  the load balancing for this PIC code over a 3D mesh instead of
a 2D one.

\bibliographystyle{plain}
\bibliography{biblio}

\end{document}